\documentclass{llncs}

\usepackage[a4paper,margin={45mm}]{geometry}

\usepackage[english]{babel}

\usepackage{microtype}

\usepackage{amsmath}
\usepackage{amssymb}
\usepackage{url}

\title{LTL Store: Repository of LTL formulae from literature and case studies
\thanks{This research was partially supported by the German Research Foundation (DFG) project KR 4890/1 \emph{Verified Model Checkers} (317422601). }}

\author{Jan K\v ret\'insk\'y  \and Tobias Meggendorfer \and Salomon Sickert}

\institute{Technical University of Munich, Germany \\\email{\{ jan.kretinsky | tobias.meggendorfer | sickert \}@in.tum.de}}

\begin{document}

\maketitle
\bigskip

Automata-theoretic approach~\cite{DBLP:conf/lics/VardiW86} is a key technique for verification and synthesis of systems with linear-time specifications, such as formulae of linear temporal logic (LTL) \cite{DBLP:conf/focs/Pnueli77}.
It proceeds in two steps: first, the formula is translated into a corresponding automaton; second, the product of the system and the automaton is further analyzed.
The size of the automaton is important as it directly affects the size of the product and thus largely also the analysis time.

Theoretical analysis and establishing the complexity of newly developed algorihtms on LTL is often the first step to asses their practicality.
However, this needs to be complemented with an extensive experimental evaluation of formulae occuring in the literature and derived from case studies. 
This continuously extended technical report collects and compares commonly used formulae from the literature and provides them in a machine readable way.

The sources as well the formulas in machine readable can be found in the LTL Store repository, accessible at 
\begin{center}
	\url{https://gitlab.lrz.de/i7/ltlstore}
\end{center}
The repository gathers and categorizes various LTL formulae and monitors their usage in the literature.

The project is loosely inspired by B\"uchi Store \cite{DBLP:conf/tacas/TsayTCC11}, a collection of B\"uchi automata.
The Spec Patterns \cite{specpatterns} webpage provides some additional ways of categorizing LTL formulas and contains typical representatives of several patterns.

\bibliographystyle{alpha} 
\bibliography{refs}

\end{document}